\documentclass[aip,jcp,amsmath,reprint,onecolumn]{revtex4-1}
\usepackage{graphicx}
\usepackage[update,prepend]{epstopdf}
\usepackage{natmove}

\renewcommand\vec{\mathbf}
\newcommand{\vv}[1]{\mathbf{#1}}
\renewcommand{\d}[1]{\ensuremath{\operatorname{d}\!{#1}}}

\usepackage{tikz}
\newcommand{\tikzcircle}[2][red]{\tikz[baseline=-0.5ex]\draw[#1,radius=#2] (0,0) circle ;}
\newcommand{\tikzsquare}[2][red]{\tikz\draw[#1] (0,0) rectangle (#2,#2) ;}
\newcommand{\tikztri}[2][red]{\tikz\draw[#1] (0,0) -- (60:#2) -- (#2,0) -- cycle ;}

\begin{document}
\title{Stability of force-driven shear flows in nonequilibrium molecular simulations with periodic boundaries}

\author{Michael P. Howard}
\email{mphoward@utexas.edu}
\affiliation{McKetta Department of Chemical Engineering, University of Texas at Austin, Austin, TX 78712}

\author{Antonia Statt}
\altaffiliation{Present address: Department of Materials Science and Engineering, University of Illinois, Urbana, IL 61801}
\affiliation{Department of Chemical and Biological Engineering, Princeton University, Princeton, NJ 08544}

\author{Howard A. Stone}
\affiliation{Department of Mechanical and Aerospace Engineering, Princeton University, Princeton, NJ 08544}

\author{Thomas M. Truskett}
\affiliation{McKetta Department of Chemical Engineering, University of Texas at Austin, Austin, TX 78712}
\affiliation{Department of Physics, University of Texas at Austin, Austin, TX 78712}

\begin{abstract}
We analyze the hydrodynamic stability of force-driven parallel shear flows
in nonequilibrium molecular simulations with three-dimensional periodic boundary
conditions. We show that flows simulated in this way can be linearly unstable,
and we derive an expression for the critical Reynolds number as a function of
the geometric aspect ratio of the simulation domain. Approximate periodic extensions
of Couette and Poiseuille flows are unstable at Reynolds numbers two orders of magnitude
smaller than their aperiodic equivalents because the periodic boundaries impose
fundamentally different constraints on the flow. This instability has important
implications for simulating shear rheology and for designing nonequilibrium
simulation methods that are compatible with periodic boundary conditions.
\end{abstract}
\maketitle

\section{Introduction}
Periodic boundary conditions (PBCs) are mainstays of molecular simulations \cite{Allen:1991,Frenkel:2002},
where they are used to construct physically relevant models using material volumes
that are often computationally restricted to be many orders of magnitude smaller
than those in experiments. In standard PBCs, particles that exit the simulation
volume reenter on the opposite face and interact with the nearest periodic images of the other particles.
PBCs mitigate surface effects from the small volume, and simulated properties can
agree well with experiments while modeling remarkably few particles \cite{Alder:1957,Stillinger:1973}.
Also, PBCs play an especially important role in nonequilibrium molecular simulations of shear rheology \cite{Hoover:1983wb}.
Although wall-driven shear flows mimicking experimental rheometry can be modeled, surface effects
unduly influence these simulations because the gap height between the surfaces that can be modeled is
restricted, leading to unrealistic confinement that overemphasizes surface effects and slip compared to experiments.
To overcome this limitation, simple or oscillatory shear flow of an unbounded fluid can be mimicked using
Lees--Edwards PBCs \cite{Lees:1972}, where the images of the fully periodic simulation cell are effectively
put in relative motion at a given strain rate.
However, determining properties like the shear viscosity by this method remains challenging, particularly at
small strain rates, because the fluctuating stress must be carefully sampled.

\begin{figure}[!h]
    \includegraphics{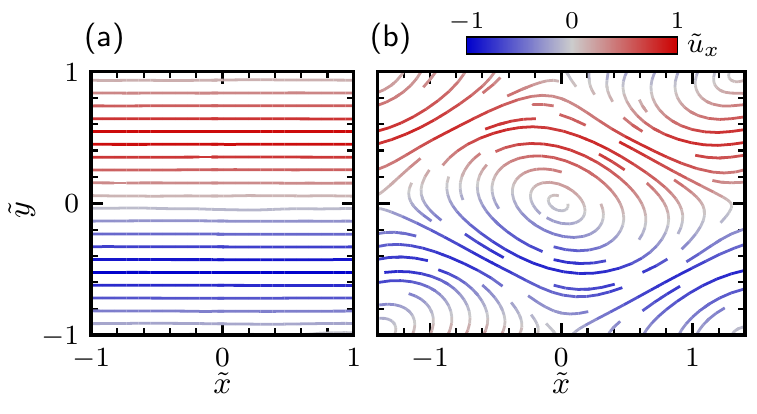}
    \caption{Streamlines for (a) stable periodic Couette flow in a cubic box at Reynolds number ${\rm Re} = 15$ and
        (b) the steady secondary flows that developed when the box was elongated to aspect ratio $\alpha = 1.4$ and
        the base flow became unstable.
        Both domains have PBCs in $x$, $y$, and $z$. The streamlines were averaged over $z$ and colored by the
        flow velocity $u_x$. All values are made dimensionless as in Eq.~\eqref{eq:orr} and with $\tilde x = x / H$.}
    \label{fig:rnesflow}
\end{figure}

Motivated by these complications, alternative simulation methods that fix the stress rather than the strain rate have been developed.
Here, we focus on one class of these approaches that includes M\"{u}ller-Plathe's reverse nonequilibrium simulation (RNES) method \cite{MuellerPlathe:1999}
and the periodic Poiseuille flow method \cite{Backer:2005}. These methods do not generate flow with forcing at the boundaries but rather
impose a spatially varying perturbation (i.e., a distributed body force) on the ``unbounded'' fluid that is compatible with the PBCs.
With thoughtful choice of the implementation and form of the perturbation, rheological properties like the shear viscosity
can be extracted from the measured flow field, which usually has rapidly converging statistics \cite{MuellerPlathe:1999}.
These methods are convenient to implement and have been widely used to simulate rheological properties of fluids such as
simple liquids \cite{Soddemann:2003,Kelkar:2007}, water \cite{Mao:2012}, ionic liquids \cite{Kelkar:2007b,Zhao:2008},
polymer solutions \cite{Nikoubashman:2017,Moghimi:2019} and melts \cite{Guo:2002,Schneider:2018,Schneider:2019}, and
colloidal dispersions \cite{Heine:2010,Cerbelaud:2017,Mountain:2017,Sambasivam:2018,OlartePlata:2018}.
However, we were recently surprised to find that the RNES method can fail
to produce the desired flow field in certain simulation geometries \cite{Statt:2019}.
RNES should generate a periodic parallel shear flow with two opposing Couette-like regions \cite{MuellerPlathe:1999}.
We obtained this flow profile (Fig.~\ref{fig:rnesflow}a) in a cubic box, but undesired steady vortices (Fig.~\ref{fig:rnesflow}b)
developed when the box was elongated in the flow direction. We found that the vortex formation
depended on the presence of PBCs in the shear-gradient direction, the aspect ratio of the domain,
and the shear rate. Moreover, the presence of vortices was not specific to the fluid model
or using the RNES algorithm to create the flow. We speculated that this might indicate a general
hydrodynamic instability underlying these nonequilibrium simulation techniques, but we
were not able to establish a relationship for the conditions under which the
instability occurred within the statistical accuracy of the simulations.

Here, we analyze the hydrodynamic stability of parallel viscous shear flows
that are generated by nonequilibrium (force-driven) simulation methods within three-dimensional PBCs.
Hydrodynamic stability is a long-studied topic \cite{Lin:1955,Drazin:2004}, but
shear flows within a fully periodic domain \cite{Meshalkin:1961,Green:1974,Grappin:1988,Thess:1992,Bena:1999,Sarris:2007,Dullin:2018}
have received less attention than wall-bounded flows. This may be partially
due to the fact that experimentally observed flows rarely, if ever, have PBCs;
however, PBCs are the norm rather than the exception in molecular simulations.
We first consider the general case of a parallel shear flow in PBCs that is driven by a
distributed piecewise continuous body force. We derive an expression for the
critical Reynolds number for the linear instability as a function of the aspect ratio of the fully periodic domain.
We then analyze the stability of three specific flows useful
for simulating shear rheology: sinusoidal (Kolmogorov) flow, periodic Poiseuille flow,
and a flow like the one generated by RNES that we call periodic Couette flow.
We perform complementary particle-based simulations that confirm the
analysis and shed valuable insight on the emergence of the instability.
For a fixed Reynolds number, the results indicate that many parallel shear flows
in PBCs become linearly unstable due to a geometric effect of the periodic domain, which
admits unstable disturbances once it becomes sufficiently long in the flow direction. This work provides a
simple framework for selecting geometries with PBCs that stabilize a flow,
requiring only the desired flow's Fourier series expansion. It also underscores
the need for caution when simulating dynamic processes with PBCs, which can
differ in unexpected ways from domains bounded by surfaces because the PBCs do
not impose the same constraints as the surfaces.

The rest of this article is organized as follows.
We perform the stability analysis in Section \ref{sec:analysis}, and then we describe technical
details of the particle-based simulations that we used to validate it in Section \ref{sec:sims}.
We compare the analysis to the simulations in Section \ref{sec:results},
showing excellent agreement between the two, before concluding in Section \ref{sec:conc}.

\section{Stability analysis} \label{sec:analysis}
In the simulation methods of interest, the fluid is modeled as discrete
interacting particles within a simulation cell having PBCs in three dimensions.
Shear flow is generated by imposing an external force that varies in one
dimension on the individual particles. For example, the periodic Poiseuille method
applies a constant-magnitude body force to all fluid particles but with a direction
that depends on whether the particles are in the upper or lower half of the simulation
cell \cite{Backer:2005}, while RNES swaps momentum between particles in two slab regions
to impose an effective average local force \cite{MuellerPlathe:1999}.
In order to analyze the hydrodynamic stability of these particle-based methods,
we formulate a continuum description of the force-driven flow that develops.

We consider the incompressible flow $\vv{u}$ of a Newtonian fluid governed by the
standard continuity and Navier--Stokes equations \cite{Deen:2012},
\begin{align}
\nabla \cdot \vec{u} &= 0, \label{eq:continuity} \\
\rho \left( \frac{\partial \vec{u}}{\partial t} + \vec{u}\cdot\nabla{\vec{u}} \right) &= -\nabla p + \mu \nabla^2\vec{u} + \vec{f},
\label{eq:momentum}
\end{align}
with density $\rho$, pressure $p$, viscosity $\mu$, and body force per volume $\vec{f}$.
The velocity field and stresses must be continuous through the periodic boundaries
at $x = \pm L$ and $y = \pm H$. The third dimension $z$ is also periodic, but its details will
not be required for our analysis.
In keeping with common practice in force-driven simulations \cite{MuellerPlathe:1999,Backer:2005},
we also require that there is no net acceleration (translation) of the entire domain due to $\vv{f}$,
i.e., its average value is zero.

Following the standard linear stability analysis procedure \cite{Lin:1955},
we assume that the velocity $\vec{u}=\vec{U}+\delta\vec{u}$ can be separated into two parts:
the steady (time-independent) base flow $\vec{U}$ and a perturbation $\delta \vec{u}$. (The pressure field,
$p = P + \delta p$ is expressed similarly.) Both $\vv{U}$ and $\vv{u}$ must satisfy the PBCs of the domain.
In particular, we are interested in the stability of parallel shear flows $\vv{U} = (U_x(y),0,0)$ generated by steady body forces $\vv{f} = (f_x(y),0,0)$, so
it suffices to consider the two-dimensional disturbance \cite{Squire:1933}, $\delta\vv{u} = e^{\omega t + i k x} \vv{v}(y)$ with $\vv{v} = (v_x, v_y, 0)$
and $\delta p = e^{\omega t + ikx} q(y)$. The specific form of $U_x$ and $f_x$ will depend on the simulation method.
Linearization of Eq.~\eqref{eq:momentum} (Appendix \ref{app:linear}) yields
\begin{equation}
\mathrm {Re} \left[ \left(\tilde \omega + i \tilde k \tilde U_x\right) \left( \frac{\d{}^2}{\d{\tilde y}^2} -\tilde k^2
\right)
-  i \tilde k  \frac{\d{}^2 \tilde U_x}{\d{\tilde y}^2} \right]  \tilde v_y =  \left( \frac{\d{}^2}{\d{\tilde y}^2} -\tilde k^2 \right)^2 \tilde v_y ,
\label{eq:orr}
\end{equation}
which has been written in dimensionless form by defining $\tilde y = y/H$ and $\tilde U_x = U_x/U$,
where $U$ is the maximum of $U_x$, which also naturally gives $\tilde v_y = v_y/U$, $\tilde k = H  k$,
and $\tilde \omega = \omega H/U$. The Reynolds number is ${\rm Re} = U H/\nu$ with $\nu=\mu/\rho$ being the kinematic viscosity.
The base flow is unstable when the real part of $\tilde \omega$ is positive for a solution $\tilde v_y$ that satisfies the PBCs in $\tilde y$.

In all but a handful of cases, Eq.~\eqref{eq:orr} must be solved numerically, e.g., by discretization or by
expanding $\tilde v_y$ in a set of orthogonal basis functions \cite{Drazin:2004}. Orszag solved Eq.~\eqref{eq:orr}
for plane Poiseuille flow using Chebyshev polynomials \cite{Orszag:1971}, but other basis functions can be used.
Due to the PBCs, it is advantageous to expand the flow fields in complex finite Fourier series that are inherently
periodic, $\tilde v_y(\tilde y) = \sum_n \tilde v_n e^{i \pi n \tilde y}$ and $\tilde U_x(\tilde y) = \sum_p \tilde U_p e^{i \pi p \tilde y}$.
These series can be directly substituted into Eq.~\eqref{eq:orr} using termwise differentiation:
\begin{align}
-\mathrm{Re} \left [ \tilde \omega \sum_n \left( (\pi n)^2 + \tilde k^2\right) \tilde v_n e^{i \pi n \tilde y}
+ i \tilde k \sum_p\sum_n \left( \pi^2(n^2-p^2) + \tilde k^2 \right) \tilde U_p \tilde v_n e^{i \pi (n+p) \tilde y} \right] \nonumber \\
= \sum_n \left((\pi n)^2 + \tilde k^2\right)^2 \tilde v_n e^{i \pi n \tilde y} . \label{eq:eigvalorig}
\end{align}
We then use the orthogonality of the basis functions to obtain
\begin{equation}
-\mathrm{Re} \left[ \tilde \omega \left( (\pi n)^2 + \tilde k^2 \right) \tilde v_n + i \tilde k \sum_p \left( \pi^2 (n^2-2np) + \tilde k^2 \right) \tilde U_p \tilde v_{n-p} \right]
= \left((\pi n)^2 + \tilde k^2\right)^2 \tilde v_n . \label{eq:eigval}
\end{equation}
Note that the sum in Eq.~\eqref{eq:eigval}, which resulted from products between $\tilde v_y$, $\tilde U_x$, and their derivatives in Eq.~\eqref{eq:orr},
is a convolution in Fourier space, as expected.

The coefficients $\tilde U_p$ are determined by the Fourier representation of a given base flow,
so Eq.~\eqref{eq:eigval} can be written as an eigenvalue problem for $\tilde v_n$ by
truncating the series to $-N \le n,p \le N$ for sufficiently large $N$. This problem
can be solved
for different values of $\tilde k$ and $\mathrm{Re}$ to find modes having unstable $\tilde \omega$.
Because  of the PBC in $x$, $\tilde k$ is restricted in the values it can take by
$k = \pi m / L$ with $m$ being an integer. In dimensionless variables, this gives $\tilde k = \pi m /\alpha$
with $\alpha = L/H$ being the aspect ratio of the domain. This constraint introduces the aspect
ratio into Eq.~\eqref{eq:eigval},
\begin{align}
    - i  \left( \frac{\pi m}{\alpha} \right) \sum_{p=-N}^N \left[1 -\frac{2 np }{(m/\alpha)^2 + n^2}\right] \tilde U_p \tilde v_{n-p}
    - \frac{\pi^2 [(m/\alpha)^2 + n^2] }{\mathrm{Re}}\tilde  v_n  =  \tilde \omega \tilde v_n . \label{eq:stability}
\end{align}
This approach can be considered a generalization of prior analysis for a sinusoidal flow \cite{Meshalkin:1961}
to an arbitrary (force-driven) parallel shear flow $U_x(y)$ in a fully periodic domain where the PBC aspect ratio $\alpha$ constrains $k$.
To find the limit of stability for a given $\alpha$, we numerically determined the eigenvalues of Eq.~\eqref{eq:stability} \cite{Orszag:1971,Dolph:1958} as
a function of Re and solved for the critical Reynolds number ${\rm Re}_{\rm c}$ that gave the least stable $\tilde \omega$ having real part $\Re(\tilde\omega) = 0$ when $m=1$.
(Disturbances with larger $m$ are unstable at larger $\alpha$ because they appear as a ratio.) In doing so,
we neglected the possibility of instabilities triggered by nonorthogonal eigenmodes \cite{Trefethen:1993}.

\section{Simulation methods} \label{sec:sims}
In order to test the stability analysis (Eq.~\eqref{eq:stability}), we simulated force-driven flows using
the computationally efficient multiparticle collision dynamics method \cite{Malevanets:1999wa,Gompper:2009is,Howard:2019}.
We used the stochastic rotation dynamics collision scheme \cite{Malevanets:1999wa}
with cubic cells of edge length $a$, fixed $130^\circ$ rotation angle \cite{Allahyarov:2002hq}, and random grid
shifting \cite{Ihle:2001ty}; a Maxwell--Boltzmann rescaling thermostat to maintain constant temperature $T$ \cite{Huang:2015fh};
and time $0.1\,\tau$ between collisions, where $\tau = a \sqrt{m/k_{\rm B}T}$, $m$ is the particle mass, and
$k_{\rm B}$ is Boltzmann's constant. The particle density was $\rho = 5\,m/a^3$, giving a liquid-like Newtonian fluid with
kinematic viscosity $\nu = 0.79\,a^2/\tau$ \cite{Ihle:2003cn,Ripoll:2005ev,Padding:2006uz}. All simulations were
performed with \textsc{hoomd-blue} (version 2.6.0) \cite{Anderson:2008vg,Glaser:2015cu,Howard:2018} using a
three-dimensional periodic simulation box with a square cross section ($H = 50\,a$ in $y$ and $z$) and $L$ varied in $x$ to span
$0.6 \le \alpha \le 3$. (The total number of particles ranged from 3 million to 15 million.)
The maximum velocity of the base shear flow was restricted to $U \lesssim 0.32\,a/\tau$ to
avoid artificial effects at large Mach numbers \cite{Lamura:2001un}, giving ${\rm Re} \le 20$.

\begin{figure}
    \includegraphics{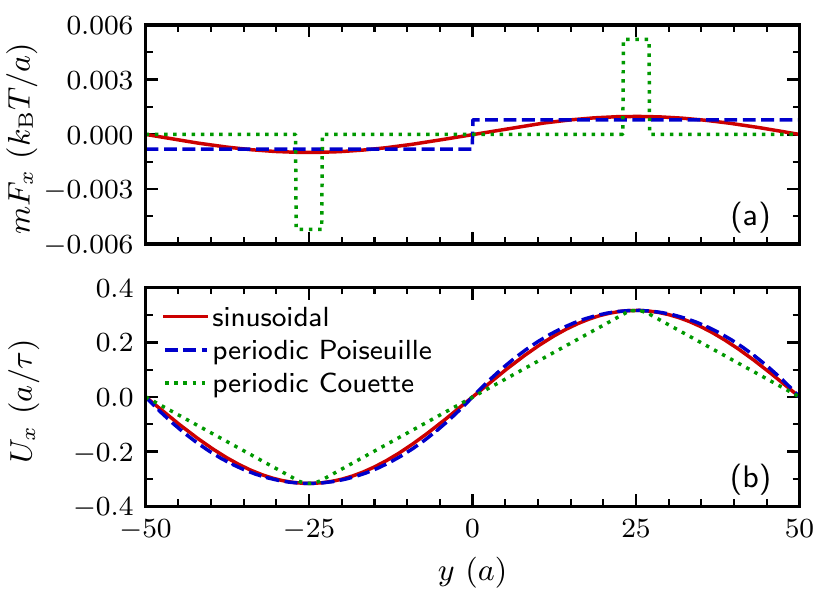}
    \caption{(a) Applied force per particle $m F_x$ and (b) resulting flow $U_x$ for the three
        cases simulated at $\rm{Re}=20$ with $H=50\,a$ (Appendix \ref{app:flow}). The maximum velocity is
        $U \approx 0.32\,a/\tau$.}
    \label{fig:plotflowandforceexample}
\end{figure}

We considered three force-driven shear flows: a sinusoidal (Kolmogorov) flow \cite{Meshalkin:1961}, a periodic Poiseuille
flow \cite{Backer:2005}, and a periodic Couette-like flow \cite{MuellerPlathe:1999,Statt:2019}. The first
is a historically well-studied problem \cite{Meshalkin:1961,Green:1974,Grappin:1988,Thess:1992,Bena:1999} that serves as a useful test of the simulations,
while flows like the latter two are commonly used to simulate shear rheology.
In our simulations, we applied a force per mass $F_x(y)$ to each particle, giving
$f_x = \rho F_x$. (All particles had unit mass $m$, so $F_x$ is also the force per particle.)
Figure~\ref{fig:plotflowandforceexample} shows $F_x$ and $U_x$ for the three flows when ${\rm Re} = 20$,
and details of their functional forms are given in Appendix \ref{app:flow}.
The periodic Poiseuille flow is the periodic extension of the classic Poiseuille flow, and it has two opposing
parabolic regions in the periodic domain. The periodic Couette-like flow, which mimics the flow created by the
RNES scheme, differs slightly from the true periodic extension of the classic Couette flow because it has
two small parabolic regions that keep the shear stress continuous and $F_x$ bounded, whereas the true periodic Couette
flow would have a step change in the shear stress. For simplicity, we will refer to the simulated periodic Couette-like
flow as periodic Couette flow.

The simulations were initialized by superimposing the base flow on the thermalized fluid and were run for $10^5\,\tau$
to reach a steady state. The velocity field $\vv{u}(x,y)$ was then measured by averaging the particle velocities in
square bins of size $a^2$ every $10\,\tau$ for $0.5 \times 10^5 \,\tau$. We used the measured $\vv{u}$ to assess whether
the targeted base flow $U_x$ was stable in the simulations. We adopted three empirical criteria to systematically classify
the base shear flow as stable if:
\begin{enumerate}
    \item The deviation of the average total kinetic energy in $y$ from equipartition
    ($\langle E_y \rangle = N_{\rm p} k_{\rm B}T/2$ for $N_{\rm p} = (\rho/m) LH^2$ particles) was less than 3 standard errors of the mean
    as computed from its own fluctuations,
    \item the absolute deviation of the flow velocity from the base flow, $|u_x(y)-U_x(y)|$, was larger than
    $10^{-3}\,a/\tau$ in fewer than $5\%$ of the bins, and
    \item the absolute value of the gradient velocity $|u_y(x)|$ was larger than $10^{-3}\,a/\tau$ in fewer
    than $5\%$ of the bins.
\end{enumerate}
If all three criteria were not met, the base flow was classified as unstable; all other cases were
unclassified. Criterion (1) is a coarse check for the development of flows directed along $y$, which are
not present in the base parallel shear flow, while criteria (2) and (3) are more detailed tests of
the measured flow against the expected flow. The thresholds for criteria (2) and (3) were chosen to
be tolerant of the numerical and statistical fluctuations in equilibrium simulations (no flow) and so base
flows with small $\alpha$ that were visibly stable were properly classified. Criterion (3) tended to be
very stringent and often failed even when criteria (1) and (2) held.

\section{Results and discussion} \label{sec:results}
\begin{figure*}[t]
    \includegraphics{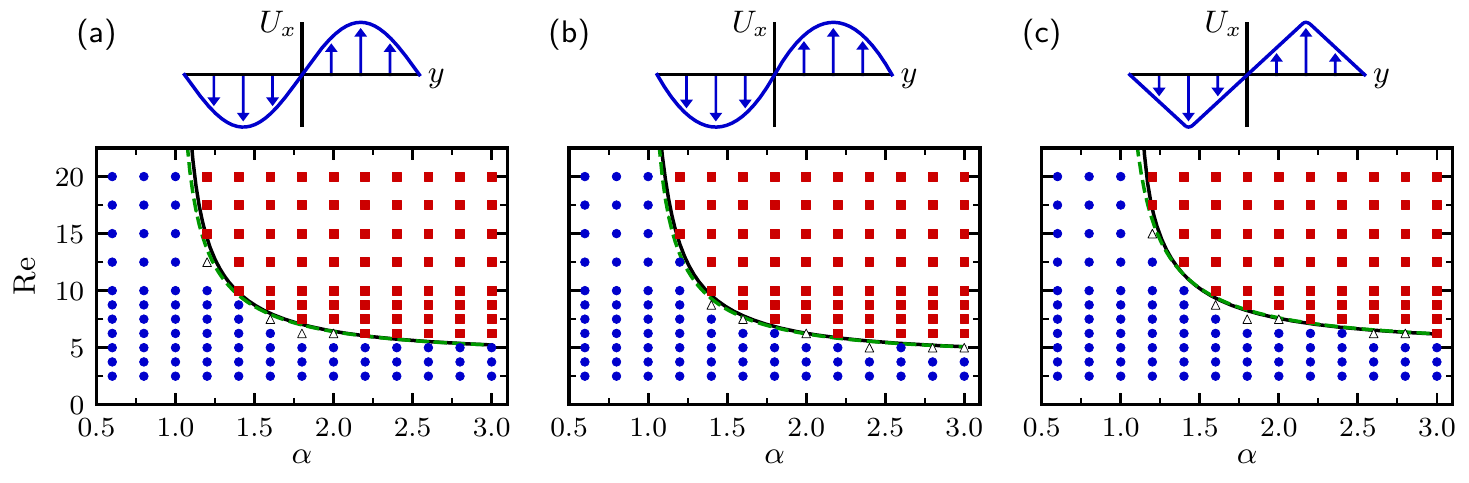}
    \caption{(top) Shear flows $U_x(y)$ and (bottom) stability diagrams in aspect ratio $\alpha = L/H$ and Reynolds number ${\rm Re} = UH/\nu$
        for the (a) sinusoidal, (b) periodic Poiseuille, and (c) periodic Couette flows. The solid black lines indicate
        the critical Reynolds number ${\rm Re}_{\rm c}$ computed from Eq.~\eqref{eq:stability}, while the dashed green lines are the three-mode
        approximation of Eq.~\eqref{eq:threemode}. The symbols show the observations from
        the simulations: stable (\tikzcircle[blue,fill=blue]{2pt}), unstable (\tikzsquare[red,fill=red]{3.5pt}),
        or unclassified (\tikztri[black]{4pt}).}
    \label{fig:stability}
\end{figure*}

We first considered the stability of the sinusoidal shear flow, $\tilde U_x(\tilde y) = \sin(\pi \tilde y)$ (Fig.~\ref{fig:stability}a).
Its Fourier coefficients are $\tilde U_{\pm 1} = \mp i/2$ and zero otherwise. Secondary flows like those shown in Fig.~\ref{fig:rnesflow}
readily formed in the simulations at sufficiently large Re and $\alpha$. During the accessible simulation time,
the flows remained stationary, although subsequent transitions to chaotic motion have been
reported \cite{Green:1974,Grappin:1988}. The bottom panel of Fig.~\ref{fig:stability}a compares the predictions of the linear
stability analysis with the simulation observations, which are in excellent agreement. Both indicate that the base flow remains
stable for all $\alpha \le 1$ over this range of Re, as proved theoretically by Meshalkin and
Sinai \cite{Meshalkin:1961}. However, Fig.~\ref{fig:stability}a also shows that some flows in domains
with $\alpha > 1$ are also stable for sufficiently small Re, and over the range of $\alpha$ tested, all flows
having ${\rm Re} \lesssim 5$ were stable. We confirmed that $\alpha$ and Re are the appropriate
dimensionless groups to use for comparison between the simulations and the stability
analysis by performing selected simulations with smaller boxes and/or higher fluid
viscosity. The flows were completely consistently with the boundaries in Fig.~\ref{fig:stability}a
when interpreted using $\alpha$ and Re.

We subsequently tested two flows commonly used in nonequilibrium molecular simulations: the periodic
extension of plane Poiseuille flow (Fig.~\ref{fig:stability}b) \cite{Backer:2005} and an approximate periodic extension
of plane Couette flow (Fig.~\ref{fig:stability}c) that mimics the RNES flow \cite{MuellerPlathe:1999}. Both were generated using constant
forces in blocks of dimensionless half-width $\tilde d = d/H$, where $\tilde d = 1/2$ for periodic Poiseuille flow
and we chose $\tilde d = 1/25$ to approximate periodic Couette flow (see Fig.~\ref{fig:plotflowandforceexample}a and Appendix \ref{app:flow}).
The corresponding Fourier coefficients are
\begin{equation}
\tilde U_p = \frac{4}{i} \frac{\sin(p \pi/2) \sin(p \pi \tilde d)}{\tilde d(1-\tilde d)(p \pi)^3} .
\end{equation}
Because the block forces are only piecewise continuous, $\tilde U_p \sim 1/p^3$ and the Fourier series for $\d{}^2 \tilde U_x/\d{\tilde y}^2$
converges pointwise rather than uniformly. We accordingly took $N = 100$ and confirmed that using $N=200$
did not significantly affect the computed critical Reynolds numbers.

Similar to the sinusoidal flow, the simulated periodic Poiseuille and Couette flows became unstable at sufficiently
large Re and $\alpha$, in excellent agreement with the stability analysis (Figs.~\ref{fig:stability}b-c).
In our previous study using RNES \cite{MuellerPlathe:1999} to create periodic Couette flow, we found stable
shear flows for $\alpha \lesssim 1.25$ \cite{Statt:2019}; this is in complete agreement with our new analysis for the studied parameters.
We are unaware of prior reports of an instability in the periodic Poiseuille flow, but its stability was
highly similar to the sinusoidal flow. This may not be surprising given that the two flows closely
resemble each other (Figs.~\ref{fig:stability}a-b). The periodic Couette flow (Fig.~\ref{fig:stability}c)
was stable over a larger parameter space than the other two flows  (see below).

Given the similar $\alpha$ dependence of ${\rm Re}_{\rm c}$ for the three flows, we posited that the stability might be controlled by
only the first (long wavelength) terms in the series expansions and applied a three-mode approximation ($N=1$) \cite{Green:1974,Bena:1999}.
This drastic simplification allows the eigenvalues of Eq.~\eqref{eq:stability}
to be computed analytically:
\begin{align}
    \tilde \omega_0 &= - \frac{\pi^2(1+\alpha^{-2})}{{\rm Re}}, \\
    \tilde \omega_\pm &= -\frac{\pi^2}{2\,{\rm Re}}\left(1 + 2\alpha^{-2} \pm \sqrt{1+\frac{8\,{\rm Re}^2}{\pi^2}\frac{1-\alpha^{-2}}{1+\alpha^2} \tilde U_{-1} \tilde U_{1}} \right).
\end{align}
$\tilde \omega_0$ and $\tilde \omega_+$ will always be negative because ${\rm Re} > 0$, but $\tilde \omega_-$ can be zero or positive.
We note that $\tilde U_1$ and $\tilde U_{-1}$ are complex conjugates, $|\tilde U_1|^2 = \tilde U_1 \tilde{U}_{-1}$,
because $\tilde U_x$ is real valued.
Solving for $\tilde \omega_- = 0$ gives a relationship between the critical Reynolds number ${\rm Re}_{\rm c}$,
$\alpha$, and $|\tilde U_1|$,
\begin{equation}
{\rm Re}_{\rm c} \approx \frac{\pi}{|\tilde U_1| \sqrt{\strut 2}} \frac{1+\alpha^{-2}}{\sqrt{\strut 1-\alpha^{-2}}}. \label{eq:threemode}
\end{equation}
There is a minimum critical Reynolds number as $\alpha \to \infty$;
this result is well-known for the sinusoidal flow \cite{Meshalkin:1961,Green:1974,Bena:1999}, where all flows having
${\rm Re} < \pi \sqrt{2} \approx 4.4$ are stable. However, ${\rm Re}_{\rm c}$ increases as $\alpha$ decreases toward one,
expanding the range of Re for which flows are stable. Both are in good agreement with the simulations, and they clarify
a point of uncertainty in Ref.~\citenum{Statt:2019} about whether there is a flow rate (Reynolds number)
below which vortices should not be observed in a simulation box of fixed size.

Equation \eqref{eq:threemode} approximates
the solution of Eq.~\eqref{eq:stability} well for all three flows when $\alpha \gtrsim 1.5$ (Fig.~\ref{fig:stability}).
This explains the similar stability of the periodic Poiseuille ($|\tilde U_1| = 16/\pi^3 \approx 0.52$) and sinusoidal ($|\tilde U_1| = 0.5$) flows, as their first Fourier modes
are nearly identical, and the stability of the periodic Couette flow to larger Re ($|\tilde U_1| \to 4/\pi^2 \approx 0.41$ as $d \to 0$). However, there are some discrepancies between Eqs.~\eqref{eq:stability} and \eqref{eq:threemode}
at smaller $\alpha$. In fact, Eq.~\eqref{eq:stability} predicts that periodic Poiseuille flow becomes unstable for ${\rm Re} \gtrsim 50$
when $\alpha = 1$, but Eq.~\eqref{eq:threemode} diverges as $\alpha \to 1$. Certain shear flows may still
be unstable in cubic domains when PBCs are used. Nevertheless, Eq.~\eqref{eq:threemode} provides a useful
estimate for either (1) choosing a Reynolds number for a fixed aspect ratio or (2) choosing
an aspect ratio for a fixed Reynolds number that keeps the base flow stable.
It also highlights the generality of the instability, as Eq.~\eqref{eq:threemode} indicates similarly
shaped flows (to lowest order) become unstable under similar conditions.

\begin{figure}
    \includegraphics{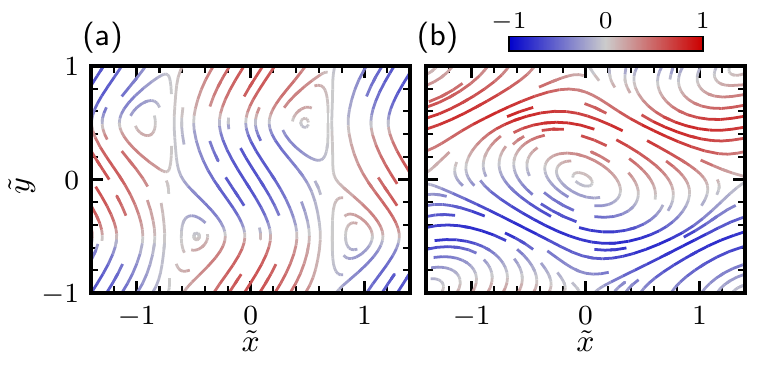}
    \caption{Streamlines for (a) the most unstable eigenmode $\vv{w}$ of the periodic Couette flow of Fig.~\ref{fig:rnesflow}b
             ($\alpha = 1.4$, ${\rm Re} = 15$) computed using Eq.~\eqref{eq:stability}
             and (b) the best-fit linear combination of $\vv{U}$ and $\vv{w}$ to the simulated $\vv{u}$.
             All quantities have been made dimensionless as in Fig.~\ref{fig:rnesflow}.}
    \label{fig:eigenmode}
\end{figure}

Most of the secondary flows that developed in our simulations had two stationary
vortices (Fig.~\ref{fig:rnesflow}b), although at the largest aspect ratio and Reynolds numbers simulated---$\alpha = 3$ and
${\rm Re} = 17.5$ and $20$---we found four stationary vortices for the sinusoidal and perioidic Poiseuille flows.
We hypothesized that the structure of some of the secondary flows might be connected to the most unstable eigenmode.
This eigenmode, $\vv{w} = e^{ikx} \vv{v}(y)$, can be reconstructed using the Fourier coefficients
for $v_y$ that comprise the eigenvector having the largest $\Re(\omega)$ and computing $v_x = (i/k) \d{v_y}/\d{y}$ based
on the incompressibility of $\vv{u}$ (Appendix~\ref{app:linear}). Figure \ref{fig:eigenmode}a shows $\vv{w}$ for the
periodic Couette flow at ${\rm Re} = 15$ and $\alpha = 1.4$, which develops into the flow shown in
Fig.~\ref{fig:rnesflow}b. It has two pairs of counterrotating vortices with streamlines primarily directed
along the shear gradient $y$.

The final steady flow need not possess the same structure as either the base flow or the most unstable eigenmode. However, we
considered as an ansatz that in some cases the simulated flow field might be well-approximated by a linear combination of the two,
$\vv{u} \approx c_1 \vv{U} + c_2 \vv{w}$, subject to a shift of the coordinates with respect to the periodic
boundaries. As an example, we performed a least-squares regression to the simulated $\vv{u}$ for the conditions
in Fig.~\ref{fig:eigenmode}, determining optimal coefficients
$c_1 = 0.765$ and $c_2 = 0.373$. The fitted flow field (Fig.~\ref{fig:eigenmode}b) bears striking similarity to the
simulated flow (Fig.~\ref{fig:rnesflow}b), having a dimensionless root-mean-squared error of $0.02$ per velocity component.

\begin{figure}
    \includegraphics{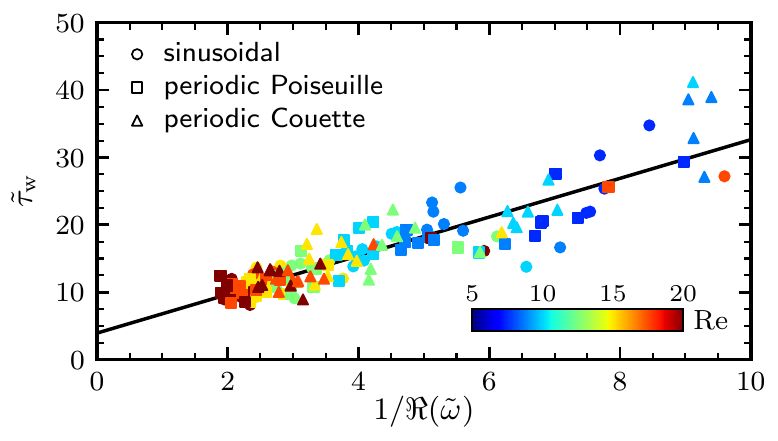}
    \caption{Dimensionless waiting time $\tilde \tau_{\rm w} = \tau_{\rm w}U/H$ for secondary flows to emerge in the simulations compared to
             the dimensionless timescale associated with the most unstable eigenmode, $1/\Re(\tilde \omega)$, for the
             sinusoidal (\tikzcircle[black,fill=white]{2pt}),
             periodic Poiseuille (\tikzsquare[black,fill=white]{3.5pt}),
             and periodic Couette flows (\tikztri[black,fill=white]{4pt}). The points are colored according to Re,
             and the line is a fit to the data with slope $2.9$.}
    \label{fig:timescale}
\end{figure}

We also noted that the waiting time $\tau_{\rm w}$ for the secondary flows to emerge in the simulations was
shorter for points farther from the curve ${\rm Re}_{\rm c}(\alpha)$. This is qualitatively expected because these conditions are
less stable based on their eigenvalues and should require smaller fluctuations (and less time) to depart from the base flow.
To quantify this, we computed $\tau_{\rm w}$ for the unstable flows by empirically fitting $E_y$ to a
hyperbolic tangent during the first $10^5\,\tau$ simulated; we defined $\tau_{\rm w}$ as the first time that $E_y$ increased by 10\% of the
difference between its initial and final values (Appendix \ref{app:timescale}). These times can be compared to $1/\Re(\omega)$,
which is the typical timescale associated with growth of the most unstable eigenmode. We found that $\tau_{\rm w}$ had
a strong linear correlation with $1/\Re(\omega)$ for all three flows (Fig.~\ref{fig:timescale}).
Moreover, we noted that the instability could take a surprisingly long
time to emerge, up to nearly $1.5 \times 10^4\,\tau$ in dimensional units. Having this estimate of the timescale for
instability is an added benefit of our analysis.

\section{Conclusions} \label{sec:conc}
The stability of force-driven shear flows simulated in fully periodic domains seems to stand in stark contrast to those in bounded ones.
Plane Couette flow is linearly stable for all infinitesimal disturbances \cite{Romanov:1973}, while plane Poiseuille
flow is linearly stable up to Re = 5772 \cite{Orszag:1971}; in practice, both become unstable in the range of
${\rm Re} \approx 10^2$ to $10^3$ \cite{Orszag:1980,Bayly:1988,Lundbladh:1991}. However, their periodic extensions (Figs.~\ref{fig:stability}b-c),
driven by distributed body forces,
are unstable at Re up to two orders of magnitude smaller. The change of stability in PBCs is not a simple
consequence of the base flow; we previously showed that an unstable periodic Couette flow
could be made stable in particle-based simulations by introducing a no-penetration boundary condition at either a half
period or a full period of the flow \cite{Statt:2019}. Instead, the PBCs impose fundamentally different
constraints: (1) the velocities and stresses are not obligated to take a certain value at a surface, expanding
the types of flows that can be realized, and (2) the PBCs restrict the wavelengths of disturbances to
those commensurate with the domain, introducing a strong geometric dependence once the domain
admits unstable modes.

We have given a simple recipe for determining the stability of parallel shear flows
in spatially periodic domains that requires only the Fourier series expansion of the flows,
and we have applied it to understand the flows we observed in nonequilibrium molecular simulations.
This recipe can be used to design well-behaved models and simulation methods and
to choose appropriate simulation parameters; in the case studied, we encourage
performing simulations in domains having $\alpha < 1$. The presence of a hydrodynamic instability in
nonequilibrium simulation methods like RNES, which we and others had not previously appreciated,
also highlights the need for caution when using PBCs to simulate certain dynamic processes, as the PBCs
may fundamentally alter the underlying physics in unexpected ways.

\begin{acknowledgments}
We thank Arash Nikoubashman and Zachary Sherman for helpful comments on this manuscript.
M.P.H. and T.M.T. acknowledge support from the Welch Foundation (Grant No.~F-1696). A.S. was supported by
the Princeton Center for Complex Materials (PCCM), a U.S. National Science Foundation Materials Research Science
and Engineering Center (Grant No.~DMR-1420541). The simulations were performed using computational resources supported
by the Princeton Institute for Computational Science and Engineering (PICSciE) and the Office of Information
Technology's High Performance Computing Center and Visualization Laboratory at Princeton University.
\end{acknowledgments}

\section*{Data Availability}
The data that support the findings of this study are openly available in
Princeton University's DataSpace at http://arks.princeton.edu/ark:/88435/dsp01zw12z8154.\cite{Statt:data}

\appendix
\section{Linearization} \label{app:linear}
Although the derivation of Eq.~\eqref{eq:orr} is widely documented \cite{Lin:1955,Drazin:2004}, we include details here for completeness.
Substituting $\vv{u}$ and $p$ into  Eqs.~\eqref{eq:continuity}
and \eqref{eq:momentum}, using that $\vec{U}$ must also be a solution
of the same, and neglecting terms of $\mathcal{O}(\delta \vec{u}^2)$ yields the linearized equations:
\begin{align}
\nabla \cdot \delta \vec{u} &= 0 \label{eq:contlin}, \\
\rho \left[\frac{\partial \delta\vec{u}}{\partial t} + \vec{U}\!\cdot\!\nabla{\delta\vec{u}}
+\delta\vec{u}\!\cdot\!\nabla{\vec{U}}  \right] &= -\nabla \delta p + \mu \nabla^2\delta\vec{u} .\label{eq:ns}
\end{align}
Eq.~\eqref{eq:contlin} gives a simple relation between $v_x$ and $v_y$,
\begin{equation}
i k v_x + v_y' = 0 \label{eq:w3} ,
\end{equation}
where the prime denotes the ordinary derivative, $v_y' = \d{v_y}/\d{y}$, while Eq.~\eqref{eq:ns} gives
\begin{align}
    \rho (\omega v_x + i k U_x v_x + U_x' v_y) &=-ikq +\mu \left[ v''_x - k^2 v_x\right] \label{eq:w1} \\
    \rho (\omega v_y + i k U_x v_y                 ) &= -q' + \mu \left[ v''_y - k^2 v_y\right] .\label{eq:w2}
\end{align}
Eq.~\eqref{eq:w3} can be inserted in Eq.~\eqref{eq:w1} and solved for $q$:
\begin{equation}
-q = \frac{\rho}{k^2}(\omega + ik U_x)v_y' + \frac{\rho}{ik} U'_x v_y - \frac{\mu}{k^2} \left[ v'''_y -
k^2 v_y'\right] .\label{eq:w4}
\end{equation}
Substituting Eq.~\eqref{eq:w4} into Eq.~\eqref{eq:w2} results in Eq.~\eqref{eq:orr}, which
we note is the well-known Orr--Sommerfeld equation \cite{Drazin:2004} after replacing $\omega$ by $i \omega$.

\section{Flow fields}\label{app:flow}
In this appendix, we give explicit functional forms for the studied periodic flow fields and the body
forces that we applied to generate them. All quantities have units consistent with the
multiparticle collision dynamics simulations (see Section \ref{sec:sims}).

The sinusoidal flow,
\begin{equation}
U_x(y)=U \sin\left(\frac{\pi y}{H}\right) ,
\end{equation}
was generated by a sinusoidal force applied to each particle based on its position,
\begin{align}
    F_x(y) = \frac{U\nu}{\left(H/\pi\right)^2}\sin\left(\frac{\pi y}{H}\right) = F \sin\left(\frac{\pi y}{H}\right).
\end{align}
We chose $U$ to obtain a certain Re, setting the force amplitude $F = (\pi\nu)^2{\rm{Re}}/H^3$.

Periodic extensions of plane Poiseuille flow and plane Couette flow were generated similarly using piecewise constant forces.
This functional form ensured that velocities and stresses in the fluid were continuous.
We defined two blocks centered at $y = \pm H/2$  of half-width $d \le H/2$ each. Then, a positive, constant force $F$
was applied to all particles in the upper block and a negative, constant force $-F$ was
applied to all particles in the lower block:
\begin{align}
    F_x(y) =
    \begin{cases}
        -F ,  &|y + H/2|\ge d\\
        F ,  &|y - H/2|\ge d\\
        0 , &\rm{otherwise}\\
    \end{cases} .
\end{align}
Linear momentum was conserved on average, but there were instantaneous fluctuations due to the distribution of particles.
By solving Eq.~\eqref{eq:momentum} piecewise using the symmetries of the problem, the flow field can be deduced:
\begin{equation}
U_x(y)/U =
\begin{cases}
-g_1(H+y), & y \le -H/2 - d \\
-g_2(-y), & |y + H/2| < d \\
g_1(y), & |y| \le H/2 - d \\
g_2(y), & |y - H/2| < d \\
g_1(H-y), & y \ge H/2 + d
\end{cases}
\end{equation}
with
\begin{align}
    g_1(y) &= \frac{2 y/H}{1-d/H}\\
    g_2(y) &= 1 - \frac{(1-2y/H)^2}{1-(1-2d/H)^2}.
\end{align}
$g_1$ is the usual linear (Couette-like) flow, and $g_2$ is the quadratic (Poiseuille-like) flow.
The maximum velocity is
\begin{equation}
U = \frac{FH^2}{8\nu}\left[1 - (1-2d/H)^2 \right].
\end{equation}
For the periodic plane Poiseuille flow \cite{Backer:2005}, the width of the blocks was chosen to
cover the full simulation box, $d = H/2$. Only $g_2$ contributes to this flow, establishing
two opposing parabolic regions, and the force magnitude required for a given Reynolds number is $F = 8 \nu^2{\rm Re}/H^3$.

By making $d$ much smaller than $H/2$, the flow is dominated by $g_1$ and
a periodic plane Couette-like flow can be generated having two linear regimes. In practice,
$d$ must be sufficiently large
that there are enough particles in each block to reliably apply the force and achieve the targeted maximum velocity.
The force magnitude required for a given Reynolds number is $F =  8 \nu^2{\rm{Re}}/\left[ H^3(1-(1-2d/H)^2)\right]$.
In the limit $d \to 0$, the shear stress has a step change at $y = \pm H/2$, and the applied forces are necessarily delta functions.
For finite $d$, however, $F$ is bounded and the flow has two small quadratic regions that keep the shear stress
continuous. We chose $d = 2\,a$ for our boxes having $H = 50\,a$, which gave predominantly Couette-like flow fields
but also ensured the blocks contained sufficient numbers of particles. The resulting flow is highly similar
to that generated by the RNES method \cite{MuellerPlathe:1999} even though the origin of the perturbation
is different.

\section{Timescale fitting} \label{app:timescale}
\begin{figure}[!h]
    \includegraphics{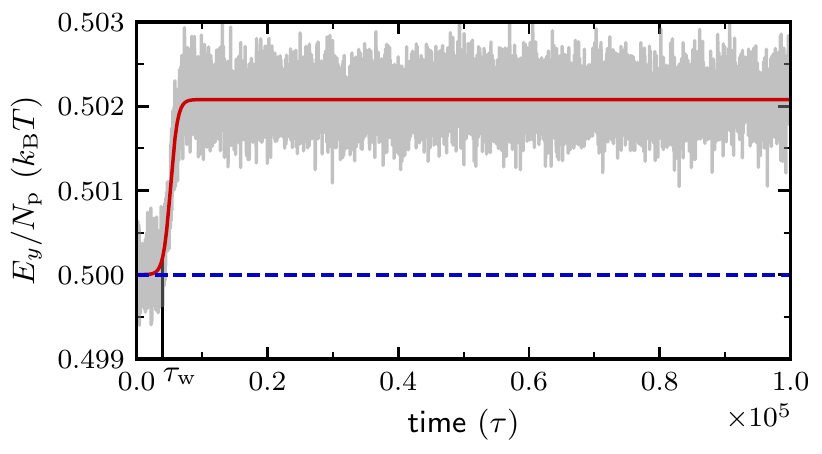}
    \caption{Fit (red) to Eq.~\eqref{eq:tanh} for kinetic energy in $y$ per particle (gray) for periodic Couette flow
        with $\alpha = 1.4$ and ${\rm Re} = 15$. The dashed blue line
        indicates the expected value of $E_y/N_{\rm p}$ from equipartition. The black line marks the
        waiting time $\tau_{\rm w}$ for occurrence of the instability.}
    \label{fig:tanh}
\end{figure}

In order to estimate the timescale for occurrence of the instability, we empirically fit the kinetic
energy $E_y$ to a hyperbolic tangent,
\begin{equation}
\frac{E_y(t)}{N_{\rm p} k_{\rm B}T} = \frac{1}{2}\left[(E_0 + E_1) + (E_1 - E_0) \tanh\left(\frac{t-\tau_0}{w}\right) \right]. \label{eq:tanh}
\end{equation}
We fixed $E_0 = 1/2$ because the system was initially thermalized, and so its kinetic energy must
obey equipartition. We then fit the final value $E_1$, the midpoint $\tau_0$, and the width $w$,
resulting in fits like those shown in Fig.~\ref{fig:tanh}.
The waiting time was defined as $\tau_{\rm w} = \tau_0 - 1.09861 w$ using the ``10--90
thickness.'' We found it challenging to reliably fit $E_y$ for a handful of conditions close to the stability
curves of Fig.~2; we accordingly neglected all points having $1/\Re(\tilde \omega) > 10$ in creating
and analyzing Fig.~4.

\bibliography{periodic_flow_stability}

\end{document}